\documentclass[letterpaper]{article}
\usepackage[preprint]{rf}
\usepackage[hyphens]{url}
\usepackage{graphicx}
\usepackage{natbib}
\usepackage{caption}
\usepackage{algorithm}
\usepackage{algorithmic}

\usepackage{newfloat}
\usepackage{listings}
\DeclareCaptionStyle{ruled}{labelfont=normalfont,labelsep=colon,strut=off}
\floatstyle{ruled}
\newfloat{listing}{tb}{lst}{}
\floatname{listing}{Listing}

\usepackage{booktabs}
\usepackage{amsmath}
\usepackage{xspace}
\usepackage{multirow}

\newif\ifhideterms
\hidetermstrue
\makeatletter
\newbox\@wmbox
\newcommand{\@rfword}{\char82\char101\char112\char114\char111\char70\char117\char115\char101}%
\newcommand{\@lcword}{\char76\char97\char121\char101\char114\char67\char97\char115\char116}%
\newcommand{\@wmark}[2]{%
  \leavevmode
  \setbox\@wmbox=\hbox{#2}%
  \edef\@wmser{\f@series}\edef\@wmbf{\bfdefault}%
  \ifx\@wmser\@wmbf \def\@wmsuf{-bf}\else \def\@wmsuf{}\fi
  \raisebox{-\dp\@wmbox}{\includegraphics[width=\wd\@wmbox,height=\dimexpr\ht\@wmbox+\dp\@wmbox\relax]{Figures/wm-#1\@wmsuf}}%
}
\newcommand{\sysname}{\ifhideterms\@wmark{rf}{\@rfword}\else\@rfword\fi\xspace}
\newcommand{\lc}{\ifhideterms\@wmark{lc}{\@lcword}\else\@lcword\fi\xspace}
\makeatother
\newcommand{\wmgfx}[2][]{\includegraphics[#1]{\ifhideterms#2-flat\else#2\fi}}

\newif\ifcombinesupplement
\combinesupplementtrue
\newcommand{\suppref}[2]{\ifcombinesupplement\ref{#1}\else#2\fi}

\title{Accelerating the Mitigation of LLM Inference Nondeterminism Across GPU Architectures}
\author{
    Liam Cooper\textsuperscript{\rm 1}, Shinnung Jeong\textsuperscript{\rm 1}, Hyeran Jeon\textsuperscript{\rm 2}, Jeffrey Young\textsuperscript{\rm 1}, Hyesoon Kim\textsuperscript{\rm 1}
}
\affiliations{
    \textsuperscript{\rm 1}Georgia Institute of Technology, \textsuperscript{\rm 2}University of California, Merced\\
}

\begin{document}

\maketitle

\begin{abstract}
Large language model (LLM) outputs are expected to be reproducible under greedy decoding, yet in practice the same model, prompt, and software stack produce different outputs on different GPUs.
The root cause is floating-point non-associativity combined with hardware-dependent kernel selection.
Inference frameworks select different matrix-multiplication kernels on each architecture, with different parallel reduction orders and unspecified tensor-core arithmetic, and the resulting rounding differences can flip output tokens.
Existing solutions have imperfect cross-architecture reproducibility and incur a significant performance penalty.
We present a solution employing a set of fixed-configuration fused-upcast GEMM kernels that load 16-bit weights from memory, upcast them to FP32 in registers, and accumulate with IEEE-754 arithmetic in a reduction order that is a pure function of the problem shape and is therefore independent of the device, its SM count, or kernel scheduling.
By fixing the floating-point reduction order as a function of problem shape alone, every GPU runs the same operation sequence, so cross-architecture reproducibility of the linear layers reduces to correct IEEE-754 arithmetic rather than to rounding differences staying below a tie-flip threshold.
We confirm our solution's linear-layer outputs are bitwise identical across NVIDIA Ampere, Ada, and Hopper GPUs, while running $1.17$ to $3.1\times$ faster end-to-end than the state-of-the-art solution and cutting weight-memory traffic in half.
\end{abstract}

\section{Introduction}

Reproducibility is a basic expectation of deployed software, and large language models are increasingly deployed inside systems that assume it.
Continuous-integration pipelines need identical regression outputs to distinguish real behavioral changes from numerical noise.
Regulated industries such as finance, healthcare, and law must be able to replay a model's exact response for auditing.
Agentic workloads rely on deterministic replay for undo/redo and debugging \citep{khatchadourian2026replayable, gundersen2018state}.
Greedy decoding (temperature~$0$, always selecting the highest-probability token, and fixing the seed) would ordinarily be expected to yield deterministic outputs.

Even with identical model weights, prompts, software versions, and greedy decoding, LLM outputs can diverge across GPU architectures, GPU counts, and batch sizes due to floating-point non-associativity and architecture-dependent tensor-core accumulations \citep{yuan2025understanding, atil2025nondeterminism, he2025nondeterminism, ieee2019, muller2018handbook}.
Low-precision formats such as BF16 further amplify these rounding differences, which can flip the argmax when the top candidate tokens are nearly tied, causing the generated sequences to diverge \citep{fasi2021numerical, sun2023dissecting, valpey2025smtformalizationmixedprecisionmatrix}.

The state-of-the-art mitigation for cross-GPU divergence is \lc \citep{yuan2025understanding}, which stores weights in BF16 but performs all computation in FP32: every linear layer upcasts its weight matrix to a temporary FP32 copy and calls the vendor GEMM.
This shrinks per-operation rounding error to the FP32 level, at which near-tie argmax flips become rare.
But \lc's design has two limitations.
First, it enforces reproducibility at the cost of performance: LLM decoding is memory-bandwidth-bound, and because the transient FP32 copy is what the GEMM actually reads, \lc's weight traffic, the quantity that sets decode latency, is fully FP32.
The BF16 storage saves resident memory while leaving memory bandwidth demand unchanged.
Second, its reproducibility is statistical rather than by construction: the vendor GEMM still selects different kernels with different reduction orders on each architecture, so different GPUs that we tested (A100, L40S, and H100) still compute different logits, and the tokens merely agree most of the time because the FP32-level differences are usually too small to flip a tie.

We present \sysname, which delivers a stronger reproducibility property than \lc at a significantly lower performance overhead.
For accelerating reproducibility, \sysname replaces the cast-then-GEMM pattern with fused-upcast GEMM kernels: BF16 weights are loaded from HBM and upcast to FP32 in registers inside the matrix-multiplication kernel, with all accumulation performed by IEEE-754 fused multiply-add (FMA) instructions whose results are bit-specified for given operands on every architecture.
Because the upcast happens entirely in registers, weight memory traffic returns to BF16 levels, halving the memory traffic during decode while eliminating the need for transient FP32 weight allocation.
To enforce a stronger reproducibility, we pin every choice that vendor libraries vary per device: kernel configurations are compile-time constants selected by problem shape alone (no autotuning), no tensor-core paths are used for the FP32 accumulation, and small-batch decode shapes use a deterministic split-$K$ scheme whose partial sums are combined in a fixed ascending order with no atomics.
The complete floating-point reduction order is therefore a pure function of the problem shape.

This design converts cross-architecture reproducibility from an empirical tendency into a verifiable kernel-level property.
We show that \sysname's linear-layer outputs are bitwise identical across three GPU architectures on every probe shape we test, spanning decode and prefill regimes, BF16 and FP32 weights, and ragged dimensions.
As a corollary of the same design, \sysname is also batch-invariant within its decode bucket: a given request's outputs are bitwise independent of how many other requests share its batch, a property identified as the key to run-to-run determinism in serving systems \citep{he2025nondeterminism} and one that \lc's vendor GEMMs do not have.

In summary, this paper makes the following contributions:
\begin{itemize}
    \item An analysis of why FP32-compute nondeterminism mitigation pipelines remain both slow and only statistically reproducible, locating the residual divergence channel in the GEMM reduction order.
    \item \sysname, a linear layer built from fixed-configuration fused-upcast Triton GEMM kernels with IEEE-754 FMA accumulation and deterministic split-$K$, whose reduction order is a pure function of problem shape.
    \item A cross-architecture bitwise validation methodology and results: \sysname's linear-layer outputs are bit-identical across Ampere, Ada, and Hopper GPUs, a property we verify with seeded probes and independent per-node hashing.
    \item An end-to-end evaluation in vLLM showing improvement in reproducibility, throughput ($1.17$--$3.1\times$), and memory (up to 1.4~GiB weight savings, converted to KV-cache capacity), over the state-of-the-art.
\end{itemize}

\section{Background and Related Work}

\subsection{Floating-Point Non-Associativity and GPU Kernels}

IEEE-754 floating-point addition rounds after every operation, so $(a+b)+c \neq a+(b+c)$ in general \citep{ieee2019, muller2018handbook}.
A dot product of length $K$ can therefore yield different results depending on how its partial sums are associated, with worst-case error growing with the accumulation depth and the condition number of the sum \citep{shanmugavelu2024impactsfloatingpointnonassociativityreproducibility}.
GPU GEMM kernels exploit this freedom aggressively: tile sizes, the number of accumulators, split-$K$ factors, and atomic reduction schemes all can change the association order, and high-performance libraries such as cuBLAS select among many such kernels using architecture- and shape-dependent heuristics.
Run-to-run determinism on a single device is typically guaranteed, but nothing constrains two different architectures to select the same kernel.

Tensor cores add a second, deeper source of architecture dependence.
The matrix-multiply-accumulate (MMA) units on NVIDIA GPUs accumulate internally with truncated (round-toward-zero) significand alignment, carry out of the standard, and per-generation differences in intermediate width and normalization.
These behaviors are officially undocumented, but are observable to cause differences across Volta, Ampere, Ada, and Hopper \citep{fasi2021numerical, sun2023dissecting, markidis2018nvidia, valpey2025smtformalizationmixedprecisionmatrix, li2024fttn, xie2025revealingfloatingpointaccumulationorders}.
Notably, ``FP32'' GEMMs on modern NVIDIA GPUs are typically executed on tensor cores in the TF32 format, which rounds inputs to a 10-bit significand \citep{nvidia_tf32_precision_format}, so even a nominally FP32 pipeline can silently inherit tensor-core arithmetic unless TF32 is explicitly disabled.\footnote{This applies at multiple layers of the stack independently: disabling TF32 in PyTorch (\texttt{torch.backends.cuda.matmul.allow\_tf32}) does not affect kernels generated by Triton, which must be constrained separately per dot operation.}
In contrast, the scalar fused multiply-add (FMA) instructions on CUDA cores implement IEEE-754 arithmetic exactly: for given operands, an FMA produces the same bits on every architecture.
This asymmetry, bit-specified scalar FMA versus unspecified MMA internals, is the foundation \sysname builds on.

\subsection{Nondeterminism in LLM Inference}

\citet{yuan2025understanding} provide the first systematic study of LLM inference divergence across GPU types, GPU counts, and batch sizes, showing that greedy decoding is not reproducible in practice: under BF16, accuracy on AIME'24 \cite{aime24} varies by up to 9\% and output lengths by up to 9{,}000 tokens across 12 runtime configurations, with reasoning models diverging on essentially 100\% of problems within the first ${\sim}100$ tokens.
Their analysis attributes the divergence to reduction-order-sensitive rounding interacting with near-tie argmax decisions: at observed divergence points the top-1/top-2 probability gap is a fraction of a percent, well within reach of BF16 rounding noise.
Their mitigation, \lc, loads weights in FP32, stores linear-layer weights in BF16, and upcasts each weight matrix back to FP32 just-in-time for its matmul, reducing divergence to below 3.4\% of problems at 34\% less memory than full FP32.

\citet{he2025nondeterminism} attribute serving nondeterminism to the lack of batch invariance: varying batch sizes change GEMM tilings and reduction orders, producing different outputs despite deterministic kernels.
Their batch-invariant BF16 kernels eliminate within-device nondeterminism at a reported ${\sim}20\%$ matmul slowdown but remain architecture-dependent.
Like their approach, \sysname{} uses batch-invariant kernels, but additionally guarantees bit-identical results across GPU architectures by restricting all arithmetic to IEEE-compliant FMAs.

Two recent systems address determinism at different layers of the stack.
LLM-42 \citep{gond2025llm42} argues that batch-invariant kernels are overly restrictive because they preclude shape-adaptive optimizations.
Instead, it restores determinism through scheduler-level verification: fast nondeterministic decoding is periodically checked against a fixed-shape deterministic replay, with rollback on mismatch, so overhead applies only to requests requiring determinism.
LLM-42 only addresses reproducibility for LLMs on the same GPU architecture, unlike \sysname{}'s goal of fast cross-architecture determinism.
Hawkeye \citep{badash2026hawkeye} reverse-engineers the architecture-specific internals of NVIDIA tensor-core MMA instructions to emulate them on CPUs, demonstrating that Ampere, Ada, and Hopper implement different MMA arithmetic.
In contrast, \sysname{} makes the dominant kernels themselves bit-identical across GPU architectures, eliminating the need for slower replay on CPU.
\sysname{} aims to reduce the verification burden by accelerating reproducibility whereas Hawkeye reports an order of magnitude slower for bitwise reproducibility.

Fusing a storage-format conversion into the GEMM has been explored by prior works.
Weight-only quantization systems dequantize INT4 weights to FP16 in registers inside the matmul for bandwidth \citep{lin2024awq, frantar2024marlin}.
\sysname takes the essence of this technique, but with the opposite numerical goal: those kernels autotune per device and accumulate on tensor cores to maximize throughput, surrendering any cross-architecture bit-for-bit agreement, whereas \sysname pins every such degree of freedom to keep the reduction order identical across devices.

Broader literature characterizes numerical variability in HPC and deep learning: compiler-induced variability \citep{Bentley_2019}, randomized testing of floating-point behavior across platforms \citep{laguna2020varity}, cross-vendor accelerator differences \citep{10820789, li2024fttn}, and the impact of tooling randomness on training \citep{zhuang2022randomness}.
Our focus is complementary: we target the inference-time, cross-architecture channel and the operations that dominate LLM FLOPs.

\section{\sysname{} Design}
\label{sec:design}
\sysname reimplements the linear layers of a hybrid-precision (BF16-storage, FP32-compute) inference pipeline as fixed-configuration Triton GEMM kernels.
Figure~\ref{fig:overview} summarizes the difference from \lc.
\lc computes each linear layer as
\begin{equation}
    y = x \, W_{\mathrm{fp32}}^\top + b, \qquad W_{\mathrm{fp32}} = \mathrm{cast}(W_{\mathrm{bf16}}),
\end{equation}
where the cast materializes a full FP32 copy of $W$ in HBM on every forward pass: one extra kernel launch, a transient allocation of size $2|W|$, and, because the GEMM reads the FP32 copy, weight traffic at FP32 width.
\sysname computes the same quantity with the upcast fused into the GEMM:
\begin{equation}
    y = x \, \mathrm{up}(W_{\mathrm{bf16}})^\top + b,
\end{equation}
where $\mathrm{up}(\cdot)$ denotes a per-tile register upcast: each BF16 weight tile is loaded from HBM, widened to FP32 in registers, multiplied against the FP32 activation tile, and accumulated in FP32.
Weights cross the memory bus at BF16 width and nothing FP32-sized is ever materialized in HBM.
Since decode-phase GEMMs are memory-bandwidth-bound and weights dominate the bytes, halving weight traffic yields lower decode latency.

The performance improvement comes without numerical compromise.
Reproducibility is supported by four design rules that constrain the principal sources of device-dependent variation:

\begin{figure}[t]
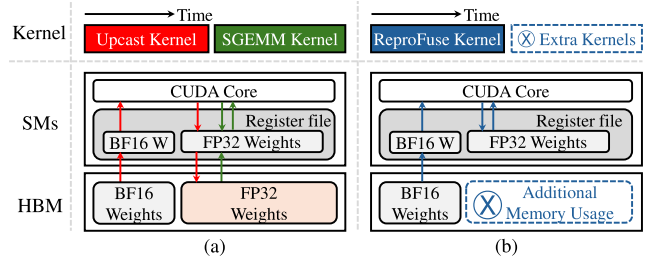

\centering
\wmgfx[width=\columnwidth]{Figures/figure1}
\caption{Dataflow in the GPU for \lc and \sysname.
(a) \lc materializes an FP32 copy of every weight matrix in HBM on every forward pass and hands it to an architecture-tuned vendor GEMM: each architecture is susceptible to reduce in a different order.
(b) \sysname loads BF16 weights directly and upcasts in registers inside a fixed-configuration IEEE-FMA kernel, reducing weight traffic and fixing the reduction order.}
\label{fig:overview}
\end{figure}

\paragraph{R1: IEEE FMA only.}
Every dot product is executed with IEEE-754 FMA on CUDA cores (\texttt{input\_precision=``ieee''} on each Triton \texttt{tl.dot}).
Triton's default for FP32 operands on Ampere and newer is TF32 tensor cores, whose internal accumulation is unspecified and architecture-dependent \citep{fasi2021numerical, nvidia_tf32_precision_format}, causing cross-architecture disagreements \cite{xie2025revealingfloatingpointaccumulationorders}.
Framework-level TF32 switches do not propagate into Triton-generated kernels, so the constraint must be applied per operation.
IEEE FMA is bit-specified for given operands on every architecture, making the elementary arithmetic outputs identical.

\paragraph{R2: No autotuning.}
Autotuners benchmark candidate configurations on the local device and keep the fastest, which makes the compiled kernel a function of the machine it was tuned on.
In our ablation testing, running the same search over the same candidate space on A100, L40S and H100 selects the same configuration for only $3$ of $9$ representative GEMM shapes, and repeating the identical search on one device changes its own answer on $1$ of $9$ shapes, because near-equal candidates trade places under measurement noise.
The split factor is a standard tuned knob that fully determines the reduction tree, and the $K$-block size sets the segment boundaries feeding it.
\sysname's configurations are instead compile-time constants selected by a pure function of the problem shape $(M, N, K)$: one for decode-shaped GEMMs ($M \le 64$) and one for prefill-shaped GEMMs, tuned once offline and pinned for every architecture.

\begin{figure}[t]
\centering
\includegraphics[width=\columnwidth]{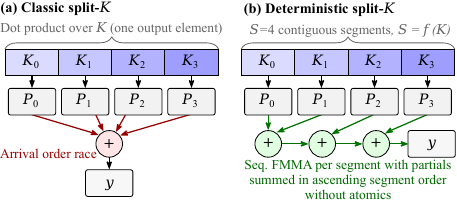}
\caption{Split-$K$ reduction of one output element.
(a) Classic split-$K$ merges partial sums with \texttt{atomicAdd}, so the reduction order is whatever order the segments' programs happen to finish in, different per run and per device.
(b) \sysname fixes the segmentation as a pure function of $K$ and sums the partials in ascending segment order in a second kernel, making the floating-point reduction tree a deterministic function of the problem shape.}
\label{fig:splitk}
\end{figure}

\paragraph{R3: Deterministic split-$\boldsymbol{K}$.}
Decode-shaped GEMMs ($M$ small, $N \times K$ large) underutilize the GPU without splitting the reduction dimension, but classic split-$K$ combines partial sums with atomic additions whose order is a scheduling race (Figure~\ref{fig:splitk}a).
\sysname splits $K$ into $S$ contiguous segments, where $S$ is a pure function of $K$; each segment is accumulated sequentially by exactly one program into a partials workspace, and a second kernel reduces the $S$ partials in fixed ascending segment order (Figure~\ref{fig:splitk}b).
No atomics appear anywhere, so the complete floating-point reduction tree is a deterministic function of shape, independent of the device, its SM count, and grid scheduling.

\paragraph{R4: Batch invariance by construction.}
Within the decode bucket, the configuration and the $K$-segmentation depend only on $K$, never on the batch dimension $M$.
A given row of the batch therefore executes the same FMA sequence at any batch size from 1 to 64: per-request outputs are bitwise independent of co-scheduled load.
This is the property \citet{he2025nondeterminism} obtain with dedicated batch-invariant kernels.

Together, R1--R4 supports reproducibility by construction: for a given input, weight, and shape, each FLOP and its position in the reduction tree are determined by the problem shape.
Cross-architecture reproducibility then no longer depends on FP32-level rounding differences staying below a tie-flip threshold; it reduces to a single premise, that each GPU implements IEEE-754 FP32 arithmetic correctly.
That premise is documented for CUDA core FP32 and confirmed by our work on A100, L40S, and H100 GPUs where the linear layer outputs are bit-identical on every shape we probe.
This is a different basis for reproducibility than \lc's, whose reduction order stays device-dependent and which can only bound the magnitude of the resulting differences.
An ablation study of R1--R4 can be found in the technical supplement (Section~\suppref{sec:supp-ablation}{C}).

We additionally found that \lc's published implementation stores the output projection (\texttt{lm\_head}) in FP32: its weight-conversion filter matches only attention and MLP projection layers, silently exempting the vocabulary projection and contradicting the stated bf16-storage policy at a cost of $\mathrm{vocab} \times \mathrm{hidden} \times 2$ bytes ($\approx$1~GiB for an 8B model with a 128k vocabulary).
We remedy this by storing the \texttt{lm\_head} in BF16 for models with untied embeddings.

\sysname covers the linear layers, which account for the majority of FLOPs and the divergence channel identified by prior analysis.
Attention, rotary embeddings, normalization, and the sampler remain the engine's FP32 implementations and may in principle select architecture-specific code paths.
The end-to-end experiments below measure whether any residual channel is observable in practice.

\section{Evaluation}
\label{sec:eval}

\subsection{Experimental Setup}

\paragraph{Hardware and software.}
We evaluate on three NVIDIA architectures: Ampere (A100), Ada Lovelace (L40S), and Hopper (H100).
All experiments run on a single GPU.
All nodes run an identical software environment: PyTorch 2.6.0 with CUDA 12.4, and Triton 3.2.0.
Following \lc, vLLM 0.8 and the xFormers attention backend is used.

\paragraph{Models and benchmarks.}
We evaluate three instruction-tuned models spanning tied and untied embeddings and a range of sizes: meta-llama/Llama-3.2-3B-Instruct \citep{meta2024llama32_3b_instruct}, Qwen/Qwen3-4B-Instruct-2507 \citep{qwen3technicalreport}, and deepseek-ai/DeepSeek-R1-Distill-Llama-8B \citep{deepseekai2025deepseekr1}.
Qwen/Qwen3-14B is added for memory profiling.
Benchmarks are GSM8K \citep{gsm8k}, MATH500 \citep{math500-2}, AIME24 \citep{aime24}, and GPQA-Diamond \citep{gpqa}.
We selected these datasets to align with \lc's evaluation and all are evaluated with greedy decoding and batch size 32.

The Llama-3.2 template normally sets a date string which gets embedded in the rendered system prompt.
Evaluation sweeps spanning a calendar day boundary therefore present different token sequences to nominally identical configurations.
This artifact can masquerade as hardware nondeterminism, manifesting as both cross-GPU and cross-seed divergence, affecting the baseline and the mitigated system alike.
We observed up to a 37.60\% difference in cross-GPU prompt divergence which would otherwise have no divergence on sweeps spanning over two calendar days.
We therefore pin the date string to a fixed value in all reported experiments.
The Qwen3-4B and DeepSeek-R1-Distill templates contain no date field and are unaffected.

\paragraph{Methods.}
We compare three inference configurations.
The unmitigated model default (BF16 storage and compute) is the baseline whose nondeterminism motivates the problem.
\sysname replaces \lc's linear layers with our kernels.
This comparison serves as our principal ablation: replacing \sysname's fixed-order kernels with architecture-tuned vendor GEMMs while holding the storage precision, compute precision, model, and surrounding inference stack fixed isolates the effect of deterministic kernel design.
The unmitigated BF16 configuration provides a complementary end-to-end ablation of the reproducibility mitigation stack.
Each (model, benchmark, GPU, method) cell is run under three different seeds.

\subsection{Cross-Architecture Determinism}
\label{sec:eval-bitwise}

We first validate the kernel-level claim directly.
A seeded probe generates fixed inputs and weights for seven GEMM cases spanning both shape buckets (decode and prefill), BF16 and FP32 weights, and ragged shapes.
The inputs are seeded on the CPU so every GPU sees identical input bits.
Across A100, L40S, and H100, all seven cases are bitwise identical under \sysname: every tensor compares equal element-for-element and every digest matches.

For the same probe through \lc's cast-then-cuBLAS path, no shape produces identical bits on all three architectures.
A100 and L40S disagree on all seven shapes, A100 and H100 on four, and H100 and L40S on five; in each divergent case 88--99\% of output elements differ, with maximum differences of $2.7\times10^{-7}$ to $4.9\times10^{-6}$ of the output scale.
The scattered pairwise agreements are cases where two architectures happened to select the same cuBLAS kernel for that shape: agreement by coincidence of undocumented kernel-selection heuristics, constrained by no specification and stable across neither shapes nor architecture pairs.
These differences sit four orders of magnitude below BF16 rounding noise, which is why \lc's tokens usually (but are not guaranteed to) agree.

The probe also corroborates batch invariance in \sysname{}.
On all three architectures, row~$0$ of a \sysname batch-$M$ GEMM is bit-identical to the batch-size-1 result for every $M \le 64$ tested.
Under \lc, row~$0$ changes bits at every tested batch size $M \ge 2$ on A100 and H100 (and at all but the smallest sizes on L40S), so \lc inherits batch-size sensitivity even at FP32 precision, while \sysname's per-request outputs are independent of co-scheduled load.

\subsection{End-to-End Determinism}
\label{sec:eval-determinism}

\begin{table*}[t]
\centering
\small
\begin{tabular}{@{}llcccccc@{}}
\toprule
& & \multicolumn{3}{c}{Cross-GPU divergence (\% of problems)} & \multicolumn{3}{c}{Cross-seed divergence (\% of pairs)} \\
\cmidrule(lr){3-5}\cmidrule(l){6-8}
Model                     & Task         & Unmitigated        & \lc     & \sysname        & Unmitigated        & \lc     & \sysname \\
\midrule
\multirow{4}{*}{DeepSeek-R1-Distill-8B} & GSM8K        &              88.86 &          0.15 & \textbf{0}      &              19.06 &             0 & \textbf{0} \\
                           & MATH500      &              93.00 &          0.40 & \textbf{0}      &              25.40 &             0 & \textbf{0} \\
                           & AIME24       &             100.00 &             0 & \textbf{0}      &                  0 &             0 & \textbf{0} \\
                           & GPQA-Diamond &             100.00 &          0.51 & \textbf{0}      &              12.74 &             0 & \textbf{0} \\
\midrule
\multirow{4}{*}{Llama-3.2-3B} & GSM8K        & 66.64    & 0  & \textbf{0}     & 17.34   & 0  & \textbf{0} \\
                           & MATH500      & 79.60    & 0  & \textbf{0}     & 13.60   & 0  & \textbf{0} \\
                           & AIME24       & 86.67    & 0  & \textbf{0} & 17.78 &        0 & \textbf{0} \\
                           & GPQA-Diamond & 30.81    & 0  & \textbf{0} & 12.07 &        0 & \textbf{0} \\
\midrule
\multirow{4}{*}{Qwen3-4B} & GSM8K        &               70.81 &          0.23 & \textbf{0.08}  &                5.96 &            0 & \textbf{0} \\
                           & MATH500      &               83.60 &             0 & \textbf{0}     &               18.33 &            0 & \textbf{0} \\
                           & AIME24       &              100.00 &             0 & \textbf{0}     &                   0 &            0 & \textbf{0} \\
                           & GPQA-Diamond &               99.49 &          0.51 & \textbf{0}     &               20.09 &            0 & \textbf{0} \\
\bottomrule
\end{tabular}
\caption{End-to-end determinism across A100, L40S, and H100.
Cross-GPU divergence: percentage of problems whose greedy token stream differs across at least one pair of GPU architectures at matched seed.
Cross-seed divergence: percentage of (problem, seed-pair) comparisons that differ on the same GPU.
Unmitigated BF16 diverges on 30.81--100\% of problems across GPUs.
\lc greatly reduces this but leaves a residual on five of twelve configurations, up to $0.51\%$.
\sysname is $\le$ \lc in every cell and is exactly $0$ wherever the divergence was driven by the linear layers.
Its one nonzero cell, Qwen3-4B GSM8K at $0.08\%$, is a single tied token localized to the unpinned attention kernel on H100 rather than the linear layers, and still improves on \lc's $0.23\%$.
The divergence metric used is further discussed in technical supplement Section~\suppref{sec:supp-divmetrics}{B}.}
\label{tab:determinism}
\end{table*}

Table~\ref{tab:determinism} reports cross-GPU and cross-seed token divergence for end-to-end inference using all three methods.
Three observations follow.

First, unmitigated inference is not remotely reproducible.
Under BF16, 31--100\% of problems (median ${\approx}85\%$) produce a different greedy token stream on at least one pair of GPU architectures, and 6--25\% of streams differ across seeds on the same GPU purely from batch-composition variation.
This is the reproducibility crisis \lc set out to fix, quantified here across three architectures.

Second, FP32 compute alone (\lc) shrinks divergence to well under 1\% on every configuration.
It does not reach zero: a residual persists on five of twelve cells, up to $0.51\%$.
\lc's guarantee is statistical: the FP32-level differences are usually too small to flip a greedy decision, but nothing forces them to zero.

Third, the controlled kernel ablation shows that \sysname eliminates the linear-layer channel entirely.
\sysname is no worse than \lc in every one of the twelve cells.
On DeepSeek it is zero across all benchmarks, and for Qwen on all but GSM8K. On Llama-3.2-3B it is zero on all four benchmarks.
The single cell where \sysname is nonzero, Qwen3-4B on GSM8K ($0.08\%$, one problem versus \lc's $0.23\%$), is a single token that sits at an exact FP32 tie on H100 while A100 and L40S resolve it with a $3.8\times10^{-6}$-nat margin.
The residual is caused by the unpinned attention kernel in layer $0$ on H100 rather than to the linear layers.
The final answer is unchanged on all three GPUs.
\sysname pins the reduction order of the linear layers, but attention, normalization, and the sampler remain vendor kernels whose reduction order stays architecture-dependent.
The pinned QKV projection is bitwise identical on all three GPUs with the first divergence being the output of the layer-$0$ attention, where the H100 departs from A100 and L40S by about $6\times10^{-8}$.
That perturbation then propagates and compounds through the residual stream, and a greedy decision flips only where it grows to exceed the local decision margin.
Such exposed positions are rare: across the $381{,}426$ decoded GSM8K positions in this benchmark only one flips.
At that position the tie is resolved by \texttt{argmax}, which deterministically returns the lower token id.
\sysname's guarantee is thus exact for the linear layers, which carry the dominant FLOPs and the divergence channel prior analysis identifies, but does not extend to the remaining non-GEMM kernels.
This confirms the Design section scope caveat and isolates fixed-order attention as the outstanding source of nondeterminism.
Across every method and cell, \sysname's cross-seed divergence is zero.

\subsection{Decision Margins and Logit Noise}
\label{sec:eval-margins}

\begin{figure}[t]
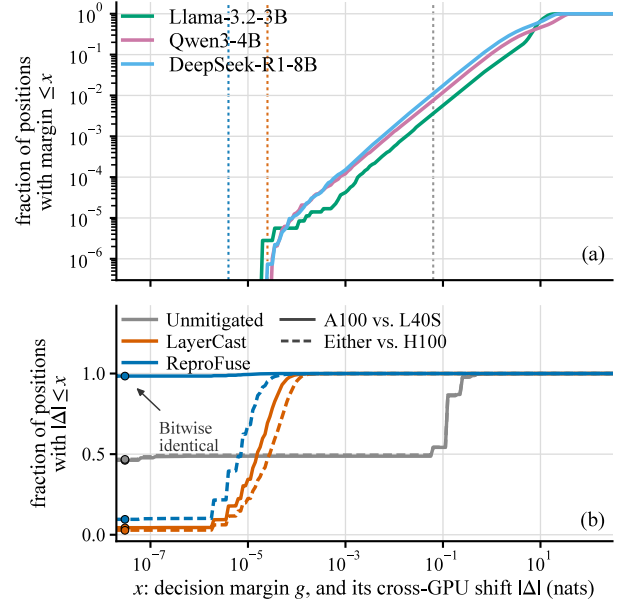

\centering
\wmgfx[width=\columnwidth]{Figures/logit-gap}
\caption{The two quantities that decide whether a token flips, plotted on one axis (nats).
(a) Cumulative distribution of the greedy decision margin $g$, the top-1/top-2 logit gap, over 2.6M decoded positions. Dotted verticals mark each method's median cross-GPU shift, so the height at which a vertical meets a model's curve is that method's per-token flip probability on that model.
(b) Cumulative distribution of the cross-GPU shift $|\Delta|$ of that same margin, split by architecture pair.
The left-hand intercept is the fraction of positions on which the two GPUs agree bitwise.}
\label{fig:logitgap}
\end{figure}

Table~\ref{tab:determinism} leaves an obvious question: why some solutions leave some model and benchmark cells reproducible and others not?
A greedy token flips between two GPUs only when the hardware perturbs the top-1/top-2 logit difference by more than that difference itself, so the answer is a comparison of two quantities measured on the same scale.
We denote the decision margin at position $t$ as $g(t) = \ell_1(t) - \ell_2(t)$, the logit gap between the best and second-best token.
The cross-GPU shift is $\Delta(t) = g_A(t) - g_B(t)$ for two architectures $A$ and $B$ at the same position.
Both are read directly from the stored top-5 log-probabilities: the log-sum-exp normalizer is token-independent and cancels in the difference, so a log-probability gap is also a logit gap.
We measure $\Delta$ only on the prefix over which the two runs still emit identical tokens, because past the first flip they are decoding different sequences and their positions are no longer comparable.
A position is exposed exactly when $g < \Delta$.

Figure~\ref{fig:logitgap}(a) shows that the models are confident at the typical position (median margin $4$ to $19$ nats) but that the low tail is heavy enough to matter: about $1\%$ of positions sit within $0.1$ nats of a tie and roughly $1$ in $10^{4}$ within $10^{-3}$.
Below $10^{-1}$ the tail is close to linear, $P(g \le x) \approx \rho x$ with $\rho$ between $0.06$ and $0.17$ per nat, which is the reason there is no safe precision short of exactness: the exposed fraction is proportional to the noise, so every order of magnitude of extra precision buys exactly one order of magnitude fewer flips and no threshold below which flips stop.

Figure~\ref{fig:logitgap}(b) places the three methods' cross-GPU shifts on that same axis.
Unmitigated BF16 shifts the margin by a median of $6.3\times10^{-2}$ nats, squarely inside the exposed tail, which is why 31 to 100\% of problems diverge.
\lc's FP32 compute moves the shift down by three and a half orders of magnitude, to a median of $1.8\times10^{-5}$ nats between A100 and L40S and $3.6\times10^{-5}$ against H100, but it almost never removes it: only $4.0\%$ and $2.3\%$ of positions respectively come out bitwise equal, because cuBLAS still reduces in a different order on each architecture.
\sysname inverts that ratio.
Between A100 and L40S its logits are bitwise identical at $98.4\%$ of 3.4M positions, and at $100.000\%$ on both Llama-3.2-3B and Qwen3-4B, with the surviving mass another order of magnitude smaller than \lc's.
Against H100 the shift is not eliminated (median $7.9\times10^{-6}$ nats, $9.6\%$ bitwise equal): the linear layers are pinned, but the attention, normalization, and sampler kernels vLLM selects on \texttt{sm90} are not.
Llama-3.2-3B adds no residual of its own.

Two checks confirm that the margin-versus-shift picture is the operative mechanism.
First, flips should occur where the margin is below the noise: at the first divergent token of every cross-GPU flip event, the median margin is $0.125$ nats under BF16 and $1.1\times10^{-5}$ to $7.4\times10^{-5}$ nats under \lc, tracking each method's own shift across four decades.
Second, the picture predicts rates, not just orderings.
Treating margin and shift as independent draws gives a per-token flip probability $p = \tfrac{1}{2}\,\mathrm{E}_{\Delta}\!\left[P(g < |\Delta|)\right]$, where the factor of one half is the single sign of the shift that moves the decision, and a per-problem divergence of $1-(1-p)^{L}$ over $L$ decoded tokens.
Across the 18 (model, benchmark, method) cells with nonzero measured divergence, spanning $0.08\%$ to $100\%$, the median ratio of predicted to measured divergence is $1.5$.
The same analysis applied within a device across seeds isolates batch invariance at the logit level: \sysname's margins are bitwise identical at $100.000\%$ of $10.0$M positions for all three models while \lc also reaches $100\%$ on all three, and unmitigated BF16 on $95$ to $99\%$.

\subsection{End-to-End Performance}
\label{sec:eval-perf}

\begin{figure*}[t]
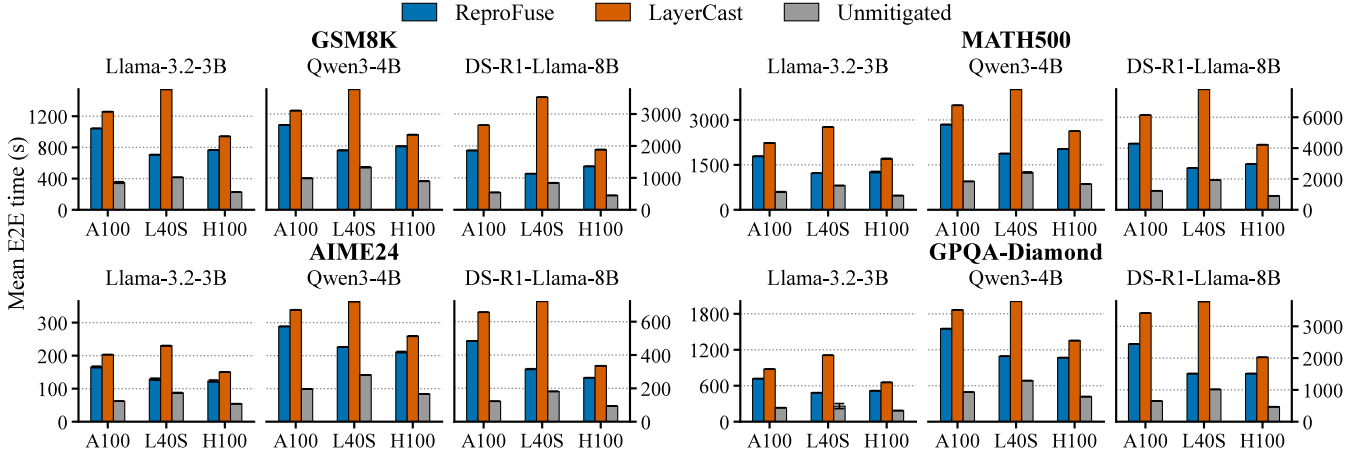

\centering
\wmgfx[width=\textwidth]{Figures/timing-comparison}
\caption{End-to-end evaluation wall-clock time (mean of 3 seeds with error bars showing the standard deviation) for \sysname, \lc, and unmitigated BF16 across three GPU architectures, three models, and four benchmarks.
\sysname is faster than \lc in every configuration, with the largest gains on the bandwidth-lean L40S, while retaining reproducibility that BF16 lacks.}
\label{fig:timing}
\end{figure*}

Figure~\ref{fig:timing} reports full-pipeline end-to-end wall-clock times.
\sysname is faster than \lc in all 36 configurations, by $1.17$--$1.43\times$ on A100 and H100 and by $1.6$--$3.1\times$ on L40S. The architecture-dependence is explained by a roofline model: \lc's decode reads FP32-width weights while \sysname reads BF16-width, so the achievable improvement grows with how bandwidth-bound the GPU is.
The A100's low FP32-compute-to-bandwidth ratio (12.5 FLOP/byte) leaves decode partially FP32-compute-bound.
The L40S sits at 106 FLOP/byte, where its \lc decode is dominated by FP32 weight reads, so \sysname reaches and, for the decode-heavy DeepSeek reasoning model, well exceeds the $2\times$ weight-traffic bound.
The bound applies to GEMM traffic alone.
End-to-end gains can exceed it because \sysname also eliminates \lc's per-layer cast kernels and transient allocations.
Measured speedups increase monotonically with the hardware ratio, exactly as the model predicts.
These are conservative, whole-workload numbers: they include prefill (where \sysname's IEEE-FMA Triton GEMM reaches $\sim$0.85$\times$ cuBLAS throughput) and CPU-side scoring.
Further analysis of the performance discrepancy between GPU architectures can be found in the technical supplement (Section~\suppref{sec:supp-l40s}{A}).

\subsection{Memory}
\label{sec:eval-memory}

\begin{table}[t]
\centering
\small
\begin{tabular}{@{}lrrr@{}}
\toprule
 & Weights (GiB) & Peak spike (GiB) & KV tokens \\
\midrule
\multicolumn{4}{@{}l}{\textbf{DeepSeek-R1-Distill-Llama-8B}} \\%
\lc & 16.92 & 0.445 & 70{,}832 \\
\sysname  & \textbf{15.94} & \textbf{0.051} & \textbf{74{,}912} \\
\midrule
\multicolumn{4}{@{}l}{\textbf{Qwen3-14B}} \\%
\lc & 30.41 & 0.673 & 128{,}288 \\
\sysname  & \textbf{28.96} & \textbf{0.060} & \textbf{133{,}536} \\
\bottomrule
\end{tabular}
\caption{Memory profile configured with vLLM's $\texttt{gpu\_memory\_utilization}=0.9$ on models with untied embeddings. ``Weights'': resident parameter bytes. ``Peak spike'': transient allocation above steady state during a forward pass. ``KV tokens'': KV-cache capacity granted by vLLM under the same memory budget, which \sysname's savings convert directly into serving capacity.}
\label{tab:memory}
\end{table}

Since total GPU memory usage is uninformative as vLLM allocates a fixed fraction for KV cache, Table~\ref{tab:memory} isolates the three memory components that differ between methods.
On untied-embedding models, \sysname's completed BF16 storage policy saves about 1--1.4~GiB of resident weights.
\lc's transient FP32 weight copies produce a per-step allocation spike the size of the largest projection matrix ($0.4$--$0.7$~GiB on these models).
\sysname's spike is smaller.
Under a fixed memory budget, vLLM converts both savings into additional KV-cache blocks, i.e., $4$--$7\%$ more KV capacity.%

\subsection{Accuracy}

\sysname changes only the order of FP32 additions relative to \lc.
Accuracy differences between the methods are numerical-noise-level.
On both Qwen3-4B tasks the scores are identical (93.93\% GSM8K, 87.2\% MATH500).
On Llama-3.2-3B the methods differ by 1--2 problems per benchmark, a difference within the numerical noise that motivates this work.
Eliminating that noise, not shifting the mean, is the contribution.
Numerical error against an FP64 reference is at the same $10^{-7}$-relative level for both methods, and \sysname's split-$K$ actually shortens accumulation chains slightly.

\section{Conclusion}

\sysname's argument covers any hardware with correctly implemented IEEE-754 FMA, which in principle could mitigate nondeterminism on various ML accelerators.
Validating bitwise portability across different vendors (where software, not just hardware, differ) is an open problem.

LLM inference diverges across GPU architectures because vendor kernels choose architecture-specific reduction orders and unspecified tensor-core arithmetic, and because low-precision rounding places many greedy decisions within flipping distance.
\lc showed that FP32 compute can make token flips more rare, but paid FP32 bandwidth for BF16 storage and left the reduction-order channel open.
\sysname closes that channel: fused-upcast GEMM kernels with IEEE-754 FMA arithmetic, shape-pure configurations, and deterministic split-$K$ make linear-layer outputs bitwise identical across Ampere, Ada, and Hopper, while running $1.17$--$3.1\times$ faster than \lc end-to-end and freeing weight memory for KV cache.
Reproducibility and efficiency are not in tension: the same discipline that fixes the reduction order also eliminates redundant memory traffic.

\bibliography{refs}

\ifcombinesupplement
  \clearpage
  \section*{Technical Supplement}
  
\appendix
\setcounter{secnumdepth}{1} \makeatletter
\renewcommand\@seccntformat[1]{\csname the#1\endcsname.\quad}
\makeatother

\section{Why the L40S Results Outperform the H100 Results under \sysname}\label{sec:supp-l40s}

On our decode-dominated workloads \sysname runs faster on the NVIDIA L40S than on the NVIDIA H100 for most layers.
This is not an anomaly.
It follows directly from the reproducibility mechanism of \sysname combined with the arithmetic profile of autoregressive decoding.
We explain it in three steps: the compute path that \sysname is restricted to, the arithmetic intensity of decode, and the resulting utilization gap between the two GPUs.

\subsection{\sysname is accelerated by FP32 vector units, not tensor cores}

To produce bitwise-identical results across GPU architectures, \sysname performs every linear-layer accumulation with IEEE-754 single-precision fused multiply-add on the CUDA (vector) cores.
It never dispatches to the tensor cores, whose accumulation order and internal rounding are unspecified and vary across GPU generations.
The hardware peak relevant to \sysname is therefore the FP32 vector throughput, not the tensor-core throughput that dominates vendor comparisons.
Table~\ref{tab:fp32peaks} lists this peak for the three GPUs we study, together with the memory bandwidth and the resulting roofline ridge point, the arithmetic intensity above which a kernel becomes compute-bound.
Two facts stand out.
First, the L40S has the highest FP32 vector throughput of the three, because each of its Ada streaming multiprocessors exposes 128 FP32 lanes at a high clock, whereas the A100 exposes 64.
Excluding the tensor cores therefore removes the H100's principal advantage.
Second, the ridge points span nearly an order of magnitude: a kernel must reach 106 FLOP/byte to be compute-bound on the L40S, but only 25.6 on the H100 and 12.5 on the A100.

\begin{table*}[t]
\centering
\small
\begin{tabular}{lrrrrr}
\toprule
GPU & FP32 vector (TFLOP/s) & FP32 lanes/SM & SMs & HBM BW (GB/s) & Ridge (FLOP/byte) \\
\midrule
A100 & 19.5 & 64  & 108 & 1555 & 12.5 \\
L40S      & 91.6 & 128 & 142 & 864  & 106.0 \\
H100 & 51.2 & 128 & 114 & 2000 & 25.6 \\
\bottomrule
\end{tabular}
\caption{FP32 vector (non-tensor-core) compute peak, memory bandwidth, and roofline ridge point for the three GPUs.
Because \sysname never uses the tensor cores, the FP32 vector peak is the relevant compute bound, and the L40S has the highest of the three.}
\label{tab:fp32peaks}
\end{table*}

\begin{table*}[t]
\centering
\small
\begin{tabular}{lrrrr}
\toprule
Projection ($N \times K$) & L40S ($\mu$s) & H100 ($\mu$s) & L40S (TFLOP/s) & H100 (TFLOP/s) \\
\midrule
qkv ($5120 \times 3072$)       & 57.2   & 70.6   & 17.6 & 14.3 \\
o   ($3072 \times 3072$)       & 56.7   & 57.2   & 10.7 & 10.6 \\
gate/up ($16384 \times 3072$)  & 191.4  & 205.1  & 16.8 & 15.7 \\
down ($3072 \times 8192$)      & 88.8   & 132.1  & 18.1 & 12.2 \\
lm head ($128256 \times 3072$) & 1701.5 & 1532.7 & 14.8 & 16.5 \\
\bottomrule
\end{tabular}
\caption{Per-layer decode latency and achieved FP32 throughput for a single Llama-3.2-3B forward at batch $M = 32$, measured per GPU under \sysname.
The L40S is faster on four of five projections.
The H100 leads only on the language-model head, the one layer wide enough to fill its SM array.}
\label{tab:decodelat}
\end{table*}

\subsection{Decode runs at low arithmetic intensity}

\begin{figure}[t]
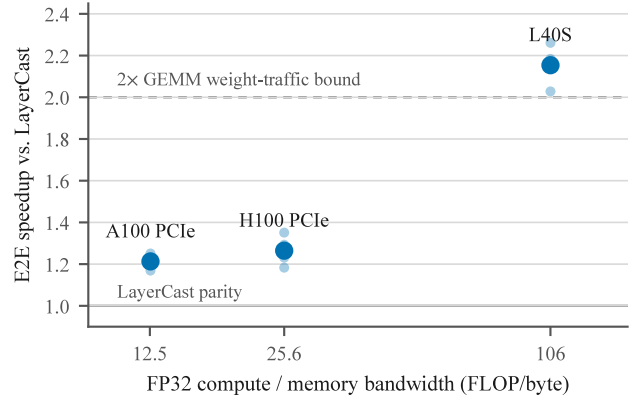

\centering
\wmgfx[width=\columnwidth]{Figures/speedup-roofline}
\caption{End-to-end speedup over \lc against each GPU's FP32-compute-to-bandwidth ratio (peak FP32 TFLOPS $\div$ memory bandwidth).
Small points are the four (model, benchmark) configurations and large points are means.}%
\label{fig:roofline}
\end{figure}

Consider a linear layer that multiplies a batch of $M$ token vectors by a weight matrix of shape $N \times K$ stored in BF16.
The layer performs $2MNK$ floating-point operations and reads $2NK$ bytes of weights, which dominate the memory traffic when $M$ is small.
Its arithmetic intensity is therefore \[ \text{AI} \;=\; \frac{2MNK}{2NK} \;=\; M \quad \text{FLOP/byte}. \] During decode, $M$ equals the number of sequences generated concurrently, which is small.
Our measurements use $M = 32$.
At this intensity the layer lies below the ridge point of every GPU in Table~\ref{tab:fp32peaks}.
In the ideal roofline limit each GPU would then be bound by weight-read bandwidth or by peak FP32, whichever is smaller.
That ideal limit favors the H100, whose bandwidth is more than twice that of the L40S. The measured ordering is the reverse, which shows that neither GPU operates near its roofline and that peak numbers alone do not determine the outcome.

\subsection{Workload Size}

The binding constraint is occupancy.
A decode GEMM at $M = 32$ produces few output tiles, and with the tensor cores disabled each tile is evaluated as a stream of scalar FP32 fused multiply-adds.
The L40S has fewer streaming multiprocessors but high per-SM FP32 throughput, so it keeps its execution units busy on these small problems and sustains a rate close to what the problem allows.
The H100 has a larger SM array and more bandwidth, but a small GEMM does not expose enough parallel work to fill it, so much of the device stays idle and its effective throughput falls well below its peak.
Table~\ref{tab:decodelat} reports per-layer decode latency and achieved FP32 throughput for a single Llama-3.2-3B forward at $M = 32$.
The L40S matches or exceeds the H100 on four of the five projections, and both GPUs realize 11 to 18 TFLOP/s.
The one exception is the language-model head, whose output dimension $N = 128256$ is large enough to generate the tiles needed to fill the H100 SM array; there the H100 is faster.
This exception supports the explanation rather than contradicting it: the H100 wins on the single layer wide enough to use it.

\subsection{Scope of the effect}

The inversion is specific to the low-intensity, low-occupancy regime of small-batch decode.
As the batch grows, the arithmetic intensity $M$ rises, the number of output tiles increases, and the H100 SM array fills.
The same holds during prefill, where $M$ spans the full prompt length.
In those regimes the H100 recovers its expected advantage.
The behavior described here should therefore be read as a property of memory- and occupancy-limited decode under a tensor-core-free reproducibility constraint, not as a general ordering of the two GPUs.

\section{Divergence Metrics and Div\_Index Results}\label{sec:supp-divmetrics}

\begin{table*}[t]
\centering
\small
\setlength{\tabcolsep}{5.5pt}
\begin{tabular}{ll cc cc cc}
\toprule
& & \multicolumn{2}{c}{Unmitigated BF16} & \multicolumn{2}{c}{\lc} & \multicolumn{2}{c}{\sysname} \\
\cmidrule(lr){3-4}\cmidrule(lr){5-6}\cmidrule(lr){7-8}
Model & Benchmark & Div\% & Div\_Index & Div\% & Div\_Index & Div\% & Div\_Index \\
\midrule
\multirow{4}{*}{Llama-3.2-3B} & GSM8K        & 66.6 & 65  & 0 & $-$ & 0 & $-$ \\
              & MATH500      & 79.6 & 106 & 0 & $-$ & 0 & $-$ \\
              & AIME24       & 86.7 & 153 & 0 & $-$ & 0 & $-$ \\
              & GPQA-Diamond & 30.8 & 114 & 0 & $-$ & 0 & $-$ \\
\midrule
\multirow{4}{*}{Qwen3-4B} & GSM8K        & 71.3 & 99   & 0.23 & 436  & 0.08 & 671  \\
                   & MATH500      & 83.4 & 204  &  0 & $-$  &  0 & $-$  \\
                   & AIME24       &  100 & 204  &  0 & $-$  &  0 & $-$  \\
                   & GPQA-Diamond & 99.5 & 131  & 0.51 & 493  &  0 & $-$  \\
\midrule
\multirow{4}{*}{DeepSeek-R1-Distill-8B} & GSM8K        & 88.9 & 104  & 0.15 & 187  &  0 & $-$  \\
                          & MATH500      & 93.0 & 189  & 0.40 & 534  &  0 & $-$  \\
                          & AIME24       &  100 & 208  &  0 & $-$  &  0 & $-$  \\
                          & GPQA-Diamond &  100 & 145  & 0.51 & 3551 &  0 & $-$  \\
\bottomrule
\end{tabular}
\caption{Divergence percentage (Div\%, the fraction of problems that fork across at least one GPU pair) alongside Div\_Index (the mean first-divergence token index over divergent cross-GPU pairs, higher is better) for all three methods.
A dash marks a cell with no divergent cross-GPU pair, which is the ideal outcome.
Div\% falls and Div\_Index rises together as numerics improve, from left to right.
For the sub-1\% \lc cells the Div\_Index mean rests on few divergent pairs (as few as one distinct problem on the GPQA-Diamond cells), so those entries are small-sample estimates.}
\label{tab:divindex}
\end{table*}

The determinism table in the main paper reports two percentages per (model, benchmark, method) cell: a cross-GPU divergence rate and a cross-seed divergence rate.
This section states precisely how those percentages are computed, and then reports the same experiments under the Div\_Index metric of \lc \citep{yuan2025understanding}, which records when a divergence first occurs rather than whether it occurs.

\subsection{How the divergence percentages are computed}

Every metric is built from one primitive: a comparison of two greedy token streams.
Fix a model, a benchmark, and a method (unmitigated BF16, \lc, or \sysname).
For a problem $p$ decoded on GPU $A$ we write the emitted greedy token stream as $y^{A}(p) = (y^{A}_0, y^{A}_1, \dots)$, taken from column $0$ of the stored top-5 token dump.
For an ordered pair of runs on GPUs $A$ and $B$ at the same seed, the first-divergence index is \[ d(p; A, B) \;=\; \min\{\, t : y^{A}_t(p) \neq y^{B}_t(p) \,\}, \] with $d = \infty$ when the two streams are identical over their common length and have equal length.
Because a single flipped token permanently forks the two sequences, $d$ is the only event that matters: everything after it is a comparison between two already-different generations.

\paragraph{Cross-GPU divergence (the Div\_Percent analog).}
Hold the seed fixed and vary the GPU.
A problem is cross-GPU divergent if $d(p; A, B) < \infty$ for at least one architecture pair $\{A, B\} \subseteq \{\text{A100}, \text{L40S}, \text{H100}\}$ at any matched seed.
The cross-GPU divergence rate is \[ \mathrm{Div\%} \;=\; \frac{\#\{\, p : p \text{ is cross-GPU divergent} \,\}}{\#\{\, p \,\}}, \] the fraction of problems whose greedy output is not identical on every architecture.
This is the per-problem quantity reported in the main paper, and it is the direct counterpart of the Div\_Percent metric of \lc, which reports the percentage of problems whose outputs differ across its 12 runtime configurations.
We aggregate at the problem level rather than the pair level because a problem that forks on any pair is not reproducible for a user, independent of how many of the pairs reproduce it.
A pair-level rate is also available in our released summary and is uniformly smaller.

\paragraph{Cross-seed divergence.}
Hold the GPU fixed and vary the seed.
The cross-seed rate is the fraction of same-GPU, same-problem seed-pair comparisons for which $d < \infty$.
Greedy decoding is seed-independent by definition, so this rate is $0$ in the ideal case, and any nonzero value is pure run-to-run nondeterminism from batch composition and scheduling rather than from the numerics of a single forward pass.
\sysname is batch-invariant by construction (rule R4 in the main paper), so its cross-seed rate is exactly $0$ in every cell.

\subsection{Div\_Index}

\lc summarizes the same forking process from the other side.
Its Div\_Index is the token index at which responses that were identical up to that point first disagree, averaged over the divergent examples of a dataset, with the convention that a higher Div\_Index means divergence is pushed later and the outputs are therefore more consistent (\lc writes $-1$ for a dataset in which nothing diverges).
We report the direct analog: the mean first-divergence index \[ \mathrm{Div\_Index} \;=\; \operatorname*{mean}_{\substack{(p, A, B)\,:\\ d(p; A, B) < \infty}} d(p; A, B), \] averaged over every divergent cross-GPU comparison at matched seed.
The two metrics answer complementary questions on the same event: Div\% counts how often a fork happens, and Div\_Index measures how deep into the generation it happens.

Table~\ref{tab:divindex} pairs the two for all twelve cells.
Three patterns hold.
First, unmitigated BF16 forks early, with a first divergence at a mean of only $65$ to $208$ tokens: on greedy reasoning traces thousands of tokens long, the architectures part ways almost at the start.
Second, FP32 compute both lowers Div\% and pushes Div\_Index far later, to as much as $3551$ tokens for \lc on DeepSeek GPQA-Diamond, which is the same statement as its low Div\% seen through the timing lens.
The surviving forks are not only rarer but also deferred.
Third, \sysname removes the linear-layer fork outright: in every cell whose divergence originates in the GEMMs there is no divergent cross-GPU pair at all, so Div\_Index is undefined (shown as a dash), the strongest possible reading of the metric.
The only finite \sysname entry is Qwen3-4B on GSM8K ($671$ tokens), a single problem whose residual originates in the unpinned attention kernel rather than the linear layers.

\section{Ablation of the Design Rules R1--R4}\label{sec:supp-ablation}

The four design rules of the main paper each forbid something that a performance-first GEMM implementation would do.
This section measures, one rule at a time, both what the rule buys and what it costs.

\paragraph{Method.}
We use a copy of the \sysname kernel in which each rule is a parameter rather than a constant, verified bitwise identical to the shipped kernel when all four rules are on.
Exactly one rule is switched off per experiment with everything else held fixed.
The test set is nine GEMM cases: the QKV, down, output and \texttt{lm\_head} projections of the evaluated models at the decode batch size of $32$, two prefill projections at a chunk of $2048$ rows, and one deliberately awkward $100 \times 1000 \times 1000$ shape that exercises the masked code paths.
Inputs are drawn from a CPU generator with a fixed seed and copied to the device, so every GPU is fed identical input bits and any later difference was introduced by the hardware or by the ablated rule.
The ragged shape is not one any of the models serves, so it is reported separately in the timing summaries.

\subsection{R1: IEEE FMA only}

Table~\ref{tab:ablation-r1} varies only the \texttt{input\_precision} argument of every \texttt{tl.dot}.

\begin{table}[t]
\centering
\small
\begin{tabular}{@{}lcccc@{}}
\toprule
Setting & Bitwise & Worst & Error & Time \\
 & all GPUs & cross-GPU & vs.\ FP64 & vs.\ R1 \\
\midrule
\texttt{ieee} (R1)   & \textbf{9/9} & -- & $4.8{\times}10^{-6}$ & $1.00$ \\
\texttt{tf32}        & 0/9 & $1.1{\times}10^{-5}$ & $7.6{\times}10^{-4}$ & $0.42$ \\
\texttt{tf32x3}      & 0/9 & $2.4{\times}10^{-7}$ & $9.3{\times}10^{-7}$ & $1.45$ \\
omitted              & 0/9 & $1.1{\times}10^{-5}$ & $7.6{\times}10^{-4}$ & $0.42$ \\
\bottomrule
\end{tabular}
\caption{R1 ablation over nine GEMM cases on A100, L40S and H100. ``Bitwise across GPUs'' counts cases whose output is identical on all three architectures.
Differences are scaled by the output magnitude.
The last column is the median decode-shape time relative to \texttt{ieee}, so values below $1$ mean the ablated setting is faster.}
\label{tab:ablation-r1}
\end{table}

Under R1 every case is bitwise identical on all three architectures.
Under TF32 tensor cores no case is.
The failure pattern is instructive: A100 and L40S agree with each other on all nine cases, and both disagree with H100 on all nine.
Ampere and Ada issue the same TF32 MMA instruction, while Hopper issues a different one, so the agreement between two of the three GPUs is a consequence of two architectures happening to share an instruction rather than of any guarantee.
This is the same coincidental-agreement pattern the main paper reports for cuBLAS kernel selection, and it is exactly what a specification-level argument is meant to replace.

\begin{table*}[t]
\centering
\small
\begin{tabular}{@{}llcccc@{}}
\toprule
& & \multicolumn{3}{c}{Configuration selected by autotuning $(B_M, B_N, B_K)$, $S$} & Bitwise across GPUs \\
\cmidrule(lr){3-5}
Shape & $(M, N, K)$ & A100 & L40S & H100 & pinned / autotuned \\
\midrule
Decode QKV, 8B          & $(32, 6144, 4096)$    & $(32,128,64)$, $8$ & $(32,128,64)$, $8$ & $(32,\textbf{256},\textbf{32})$, $8$ & yes / yes \\
Decode down, 8B         & $(32, 4096, 14336)$   & $(32,128,64)$, $8$ & $(32,128,64)$, $8$ & $(32,128,64)$, $8$ & yes / yes \\
Decode QKV, Qwen3-4B    & $(32, 6144, 2560)$    & $(32,128,64)$, $8$ & $(32,128,64)$, $8$ & $(32,\textbf{256},\textbf{32})$, $8$ & yes / yes \\
Decode down, Qwen3-4B   & $(32, 2560, 9728)$    & $(32,64,64)$, $8$  & $(32,64,64)$, $8$  & $(32,\textbf{128},64)$, $8$ & yes / yes \\
Decode output proj.\    & $(32, 4096, 4096)$    & $(32,128,32)$, $8$ & $(32,128,\textbf{64})$, $\textbf{4}$ & $(32,128,32)$, $8$ & yes / \textbf{no} \\
Decode \texttt{lm\_head}, 3B & $(32, 128256, 3072)$ & $(32,128,64)$, $2$ & $(32,128,64)$, $2$ & $(32,128,64)$, $\textbf{1}$ & yes / \textbf{no} \\
Prefill QKV, 8B         & $(2048, 6144, 4096)$  & $(128,256,32)$, $1$ & $(128,256,32)$, $1$ & $(128,256,32)$, $1$ & yes / yes \\
Prefill down, 8B        & $(2048, 4096, 14336)$ & $(128,\textbf{128},32)$, $1$ & $(128,256,32)$, $1$ & $(128,256,32)$, $1$ & yes / yes \\
Ragged probe            & $(100, 1000, 1000)$   & $(16,64,32)$, $1$  & $(16,64,32)$, $1$  & $(16,64,32)$, $1$  & yes / yes \\
\bottomrule
\end{tabular}
\caption{R2 ablation.
Each architecture runs the identical search over the identical candidate space.
The last column compares the output of the pinned configuration across the three GPUs, and then the output of each GPU's own autotuned choice across the three GPUs.}
\label{tab:ablation-r2}
\end{table*}

The accuracy column shows that the ablation also gives up what the FP32 pipeline was for.
TF32 carries ten mantissa bits, and its error against an FP64 reference is $4.0$ to $7.6 \times 10^{-4}$ of the output scale, three orders of magnitude above IEEE FMA's $4.8 \times 10^{-6}$ and of the same order as BF16 rounding.
A hybrid-precision pipeline that computes in TF32 is therefore not computing in FP32 in any sense that matters.
The three-pass \texttt{tf32x3} variant recovers the accuracy but not the reproducibility, since it is built from the same architecture-dependent MMA.

R1 is not free at the kernel level: IEEE FMA runs on the CUDA cores, so the GEMMs are $2.4\times$ slower than TF32 at the median decode shape and $5.6\times$ slower at the median prefill shape.
That cost is measured against an alternative that fails both goals of the pipeline, and it is not a cost against our baseline: \lc's cuBLAS SGEMM path also executes on the FP32 units, so the end-to-end comparison in the main paper is between two FP32 CUDA-core implementations, and \sysname is the faster of the two.

Finally, the last row substantiates the claim that framework-level switches do not reach Triton.
Every node in this experiment ran with \texttt{torch.backends.cuda.matmul.allow\_tf32 = False}.
Simply omitting the \texttt{input\_precision} argument reproduced the TF32 result bit for bit on $9/9$ cases on all three GPUs, and never matched the IEEE result.
The constraint has to be applied per operation.

\subsection{R2: No autotuning}

We replace the pinned configuration with a real autotuner: the same candidate space of $18$ tile configurations crossed with split factors $S \in \{1,2,4,8\}$, benchmarked on the local device, run identically on each architecture. we replace the pinned configuration with a benchmark-and-select search of the kind \texttt{triton.autotune} implements.
The candidate space is filtered by one shared-memory bound for all three GPUs, so a different winner reflects a different choice and not a different menu.
Table~\ref{tab:ablation-r2} lists what each device picks.

The same configuration wins on all three architectures for only $3$ of the $9$ shapes.
Autotuning is not a function of just the device.
Repeating the identical search three times on one node changed the winner on $1$ of $9$ shapes on the A100, because two near-equal candidates trade places under measurement noise.

That these choices are numerical choices, and not merely scheduling choices, is visible on a single device.
Within the decode candidate space the $72$ candidates produce $4$ distinct outputs, and on $4$ of the $27$ (shape, GPU) cells the autotuner's pick already disagrees bitwise with the pinned configuration on that same GPU.
Composed across architectures, the pinned configuration is bitwise identical on $9/9$ shapes while the autotuned one is identical on $7/9$.

One nuance is worth stating precisely, because it narrows where R2 does its work.
With R1 and R3 in force, the accumulation over the reduction dimension is strictly ascending in $k$, so $B_M$, $B_N$, the warp count and the pipeline depth are numerically inert: they change the schedule but not the reduction tree.
The numerical content of the tuner's choice therefore flows through the split factor, and through $B_K$ where it changes the segmentation.
R2 remains necessary for three reasons.
The split factor is a tuned knob in every practical Triton GEMM and is precisely the one that moves the bits; the inertness of the remaining knobs is an observed property of one Triton version's lowering rather than a specification, and pinning is what makes reproducibility independent of it.
An autotuner optimizes time, so given the choice it removes R1 as well, selecting TF32 on $26$ of the $27$ (shape, GPU) cells.

Pinning does not incur significant cost over the entire LLM pipeline.
Over the eight served shapes the pinned configuration takes $1.4\%$ longer than the best found configuration at the median and $21.9\%$ longer at worst, the worst case being the \texttt{lm\_head} shape on the L40S. The synthetic ragged probe is the one outlier, at $29$ to $70\times$: a $100 \times 1000 \times 1000$ problem gives the pinned $128 \times 128$ prefill tile only eight output tiles to spread over more than a hundred SMs.
No served prefill chunk is that small, but the case does mark the boundary of where a single pinned prefill configuration is appropriate.

\begin{table*}[!ht]
  \centering
  \small
  \caption{Hardware specifications of the evaluation machines.}
  \label{tab:hardware-specs}
  \begin{tabular}{llll}
    \toprule
    \textbf{Component} & \textbf{L40S} & \textbf{H100} & \textbf{A100} \\
    \midrule
    \multicolumn{4}{l}{\textit{GPU}} \\
    \quad Model & NVIDIA L40S PCIe (SM89) & NVIDIA H100 PCIe (SM90) & NVIDIA A100 PCIe 40GB (SM80) \\
    \quad Count & 1$\times$ & 1$\times$ & 1$\times$ \\
    \quad VRAM & 45\,GiB & 80\,GiB & 40\,GiB \\
    \midrule
    \multicolumn{4}{l}{\textit{CPU}} \\
    \quad Model & Intel(R) Xeon(R) CPU Max 9468 & Intel(R) Xeon(R) Gold 6454S & Intel(R) Xeon(R) Gold 6454S \\
    \quad Sockets & 2 & 2 & 2 \\
    \quad Cores / Threads & 96 / 192 & 64 / 128 & 64 / 64 \\
    \quad Max.\ Freq. & 2.1\,GHz & 3.4\,GHz & 3.4\,GHz \\
    \midrule
    \multicolumn{4}{l}{\textit{Memory}} \\
    \quad System RAM & 503\,GiB & 503\,GiB & 503\,GiB \\
    \bottomrule
  \end{tabular}%
\end{table*}

\subsection{R3: Deterministic split-$\boldsymbol{K}$}

R3 makes two commitments, and we ablate them separately.

\begin{table}[t]
\centering
\small
\begin{tabular}{@{}lcccc@{}}
\toprule
& \multicolumn{3}{c}{Distinct outputs, $50$ runs} & Ordered \\
\cmidrule(lr){2-4}
Decode shape & none & ordered & atomic & vs.\ none \\
\midrule
QKV, 8B          & 1 & \textbf{1} & 50 & $1.9$--$4.1\times$ \\
Down, 8B         & 1 & \textbf{1} & 50 & $2.8$--$4.4\times$ \\
QKV, Qwen3-4B    & 1 & \textbf{1} & 50 & $1.9$--$2.6\times$ \\
Down, Qwen3-4B   & 1 & \textbf{1} & 50 & $3.7$--$5.6\times$ \\
Output proj.\    & 1 & \textbf{1} & 50 & $2.1$--$3.1\times$ \\
\texttt{lm\_head}, 3B & 1 & \textbf{1} & 1 & $0.8$--$1.0\times$ \\
\bottomrule
\end{tabular}
\caption{R3 ablation on the six decode shapes, run-to-run on a fixed device.
Counts are identical on A100, L40S and H100.
The last column is the speedup of ordered split-$K$ over no split, ranged over the three GPUs.}
\label{tab:ablation-r3}
\end{table}

\paragraph{Atomics.}
Table~\ref{tab:ablation-r3} repeats each variant $50$ times on one device.
Classic split-$K$ returned $50$ different results in $50$ runs on $15$ of the $18$ (shape, GPU) cells, with run-to-run spreads up to $2.2 \times 10^{-7}$ of the output scale, the same order as the cross-architecture differences the main paper measures for \lc and therefore large enough to flip a token at a narrow decision margin.
The three exceptions are the \texttt{lm\_head} shape, where the merge order happened to be stable on all three GPUs across all $50$ runs, which is the point rather than a counterexample: the order is a race, and whether it bites is a property of the schedule and not of the program.
Ordered split-$K$ and no split each returned exactly one result in every run on every GPU, and both are bitwise identical across all three architectures on all six shapes, while atomic split-$K$ is identical across architectures on one.

Split-$K$ is not optional for the decode shapes: the ordered version is $1.9$ to $5.6\times$ faster than no split on the five projection shapes.
Only \texttt{lm\_head}, whose $N$ is already large enough to fill the machine, prefers no split.
Ordering the reduction is free: the ordered variant is $4.2\%$ faster than the atomic one at the median, since writing $S$ partials and summing them in a second pass costs no more than contending on atomics.

\paragraph{The split factor must be shape-pure.}
Sweeping $S \in \{1,2,4,8,16\}$ produces five distinct outputs on all six shapes, so the split factor fully determines the result.
A performance-first rule that sizes the split by wave quantization, which is how CUTLASS-style GEMMs choose their slices, selects $S = 2$, $8$ and $16$ on A100, L40S and H100 for the same decode QKV shape, because the three devices have $108$, $142$ and $114$ SMs.
Its output consequently differs across architectures on four of the six shapes.
Making $S$ a function of $K$ alone is what removes the SM count from the numerics.

\subsection{R4: Batch invariance by construction}

We compute row $0$ of the same input at batch sizes $M = 1$ to $64$ and compare its bits against the $M = 1$ result, under four policies: \sysname, a variant whose split factor is occupancy-sized, a variant whose configuration and split are re-searched by measurement at every batch size as a Triton autotuner keyed on $M$ would do, and cuBLAS.

\begin{table}[t]
\centering
\small
\begin{tabular}{@{}lcccc@{}}
\toprule
& \multicolumn{4}{c}{Shapes with row $0$ independent of $M$ (of 6)} \\
\cmidrule(l){2-5}
GPU & \sysname & occupancy & per-$M$ tuned & cuBLAS \\
\midrule
A100 & \textbf{6} & 1 & 2 & 0 \\
L40S & \textbf{6} & 1 & 1 & 0 \\
H100 & \textbf{6} & 2 & 1 & 0 \\
\bottomrule
\end{tabular}
\caption{R4 ablation over twelve batch sizes in $1 \le M \le 64$.
A shape counts only if row $0$ is bitwise identical at every one of the twelve.
Under cuBLAS, row $0$ differs from the $M = 1$ result at eleven of the twelve batch sizes on every shape and every GPU.}
\label{tab:ablation-r4}
\end{table}

\sysname is batch-invariant on all $18$ (shape, GPU) cells.
The occupancy-sized and per-$M$ tuned variants are invariant on $4$ of $18$ each, and cuBLAS on none, differing from the single-request result at eleven of the twelve batch sizes everywhere.
The two ablated policies also lose cross-architecture agreement: comparing row $0$ at $M = 32$ across the three GPUs, \sysname matches on all six shapes while the ablated policies match on two or fewer.
This is the same mechanism seen in R3, since a batch-dependent tile count feeds a device-dependent split factor.

R4 is cheap but not free.
Re-tuning at every batch size is $5$ to $25\%$ faster than the shape-pure policy at $M \le 16$ on the A100 and H100, $2$ to $9\%$ faster at $M = 32$, and within noise on the L40S. Buying per-request independence from co-scheduled load therefore costs a few percent of GEMM time at the batch size used throughout the paper, which the end-to-end results absorb.

R4's claim is stated for the decode bucket, $M \le 64$. row $0$ changes at $M = 65$ under every policy including \sysname, because that crosses the documented shape-bucket boundary into the prefill configuration.
Immediately above that boundary the pinned prefill configuration is poorly matched to the shape, and a tuned kernel is roughly $3\times$ faster at $M = 65$ to $128$.

\fi

\end{document}